   \documentclass[12pt]{article}
   \usepackage{latexsym}
\else\oddsidemargin25mm\evensidemargin25mm\marginparwidth25mm\fi%
\def\theequation{\arabic{section}.\arabic{equation}}

\renewcommand{\theequation}{\thesection.\arabic{equation}}
\begin{document}
\makeatletter \@addtoreset{equation}{section} \makeatother
\renewcommand{\theequation}{\thesection.\arabic{equation}}
\newcommand{\ft}[2]{{\textstyle\frac{#1}{#2}}}
\newcommand{\QED}{{\hspace*{\fill}\rule{2mm}{2mm}\linebreak}}
\def\dop{{\rm d}\hskip -1pt}
\def\bfone{\relax{\rm 1\kern-.35em 1}}
\def\bfzero{\relax{\rm I\kern-.18em 0}}
\def\inbar{\vrule height1.5ex width.4pt depth0pt}
\def\IC{\relax\,\hbox{$\inbar\kern-.3em{\rm C}$}}
\def\ID{\relax{\rm I\kern-.18em D}}
\def\IF{\relax{\rm I\kern-.18em F}}
\def\IK{\relax{\rm I\kern-.18em K}}
\def\IH{\relax{\rm I\kern-.18em H}}
\def\II{\relax{\rm I\kern-.17em I}}
\def\IN{\relax{\rm I\kern-.18em N}}
\def\IP{\relax{\rm I\kern-.18em P}}
\def\IQ{\relax\,\hbox{$\inbar\kern-.3em{\rm Q}$}}
\def\IR{\relax{\rm I\kern-.18em R}}
\def\IG{\relax\,\hbox{$\inbar\kern-.3em{\rm G}$}}
\font\cmss=cmss10 \font\cmsss=cmss10 at 7pt
\def\ZZ{\relax\ifmmode\mathchoice
{\hbox{\cmss Z\kern-.4em Z}}{\hbox{\cmss Z\kern-.4em Z}}
{\lower.9pt\hbox{\cmsss Z\kern-.4em Z}} {\lower1.2pt\hbox{\cmsss
Z\kern .4em Z}}\else{\cmss Z\kern-.4em Z}\fi}
\def\a{\alpha} \def\b{\beta} \def\d{\delta}
\def\e{\epsilon} \def\c{\gamma}
\def\G{\Gamma} \def\l{\lambda}
\def\L{\Lambda} \def\s{\sigma}
\def\cA{{\cal A}} \def\cB{{\cal B}}
\def\cC{{\cal C}} \def\cD{{\cal D}}
    \def\cF{{\cal F}} \def\cG{{\cal G}}
\def\cH{{\cal H}} \def\cI{{\cal I}}
\def\cJ{{\cal J}} \def\cK{{\cal K}}
\def\cL{{\cal L}} \def\cM{{\cal M}}
\def\cN{{\cal N}} \def\cO{{\cal O}}
\def\cP{{\cal P}} \def\cQ{{\cal Q}}
\def\cR{{\cal R}} \def\cV{{\cal V}}\def\cW{{\cal W}}
%
%
%
\def\crr{\crcr\noalign{\vskip {8.3333pt}}}
\def\tilde{\widetilde}
\def\bar{\overline}
\def\us#1{\underline{#1}}
\let\shat=\hat
\def\hat{\widehat}
\def\hyp{\vrule height 2.3pt width 2.5pt depth -1.5pt}
\def\square{\mbox{.08}{.08}}
\def\Coeff#1#2{{#1\over #2}}
\def\Coe#1.#2.{{#1\over #2}}
\def\coeff#1#2{\relax{\textstyle {#1 \over #2}}\displaystyle}
\def\coe#1.#2.{\relax{\textstyle {#1 \over #2}}\displaystyle}
\def\half{{1 \over 2}}
\def\shalf{\relax{\textstyle {1 \over 2}}\displaystyle}
\def\dag#1{#1\!\!\!/\,\,\,}
\def\to{\rightarrow}
\def\notin{\hbox{{$\in$}\kern-.51em\hbox{/}}}
\def\shdot{\!\cdot\!}
\def\ket#1{\,\big|\,#1\,\big>\,}
\def\bra#1{\,\big<\,#1\,\big|\,}
\def\equaltop#1{\mathrel{\mathop=^{#1}}}
\def\Trbel#1{\mathop{{\rm Tr}}_{#1}}
\def\inserteq#1{\noalign{\vskip-.2truecm\hbox{#1\hfil}
\vskip-.2cm}}
\def\attac#1{\Bigl\vert
{\phantom{X}\atop{{\rm\scriptstyle #1}}\phantom{X}}}
\def\exx#1{e^{{\displaystyle #1}}}
\def\del{\partial}
\def\delbar{\bar\partial}
\def\nex#1{$N\!=\!#1$}
\def\dex#1{$d\!=\!#1$}
\def\cex#1{$c\!=\!#1$}
\def\eg{{\it e.g.}} \def\ie{{\it i.e.}}
%
\def\cS{{\cal K}}
\def\IE{\relax{{\rm I\kern-.18em E}}}
\def\cE{{\cal E}}
\def\rt{{\cR^{(3)}}}
\def\IGam{\relax{{\rm I}\kern-.18em \Gamma}}
\def\IGa{\IA}
\def\LG{Lan\-dau-Ginz\-burg\ }
\def\cV{{\cal V}}
\def\Rt{{\cal R}^{(3)}}
\def\wabc{W_{abc}}
\def\WABC{W_{\a\b\c}}
\def\W{{\cal W}}
\def\tft#1{\langle\langle\,#1\,\rangle\rangle}
\def\IA{\relax{\hbox{{\rm A}\kern-.82em {\rm A}}}}
\let\picfuc=\fp
\def\hata{{\shat\a}}
\def\hatb{{\shat\b}}
\def\hatA{{\shat A}}
\def\hatB{{\shat B}}
\def\bv{{\bf V}}
\def\spg{special geometry}
\def\sc{SCFT}
\def\leel{low energy effective Lagrangian}
\def\pf{Picard--Fuchs}
\def\pfS{Picard--Fuchs system}
\def\el{effective Lagrangian}
\def\Fb{\overline{F}}
\def\nablab{\overline{\nabla}}
\def\Ub{\overline{U}}
\def\Db{\overline{D}}
\def\zb{\overline{z}}
\def\eb{\overline{e}}
\def\fb{\overline{f}}
\def\tb{\overline{t}}
\def\Xb{\overline{X}}
\def\Vb{\overline{V}}
\def\Cb{\overline{C}}
\def\Sb{\overline{S}}
\def\delb{\overline{\del}}
\def\Gammab{\overline{\Gamma}}
\def\Ab{\overline{A}}
\def\Anh{A^{\rm nh}}
\def\alphab{\bar{\alpha}}
\def\cy{Calabi--Yau}
\def\cabg{C_{\alpha\beta\gamma}}
\def\B{\Sigma}
\def\Bh{\hat \Sigma}
\def\Kh{\hat{K}}
\def\Knh{{\cal K}}
\def\A{\Lambda}
\def\Ah{\hat \Lambda}
\def\R{\hat{R}}
\def\V{{V}}
\def\T{T}
\def\Gammah{\hat{\Gamma}}
\def\twot{$(2,2)$}
\def\K{K\"ahler}
\def\rat{({\theta_2 \over \theta_1})}
\def\lv{{\bf \omega}}
\def\w{w}
\def\CP{C\!P}
\def\o#1#2{{{#1}\over{#2}}}
\newcommand{\be}{\begin{equation}}
\newcommand{\ee}{\end{equation}}
\newcommand{\ba}{\begin{eqnarray}}
\newcommand{\ea}{\end{eqnarray}}
\newtheorem{definizione}{Definition}[section]
\newcommand{\bd}{\begin{definizione}}
\newcommand{\ed}{\end{definizione}}
\newtheorem{teorema}{Theorem}[section]
\newcommand{\bth}{\begin{teorema}}
\newcommand{\eth}{\end{teorema}}
\newtheorem{lemma}{Lemma}[section]
\newcommand{\blem}{\begin{lemma}}
\newcommand{\elem}{\end{lemma}}
\newcommand{\brr}{\begin{array}}
\newcommand{\err}{\end{array}}
\newcommand{\nn}{\nonumber}
\newtheorem{corollario}{Corollary}[section]
\newcommand{\bcorol}{\begin{corollario}}
\newcommand{\ecorol}{\end{corollario}}
\def\twomat#1#2#3#4{\left(\begin{array}{cc}
 {#1}&{#2}\\ {#3}&{#4}\\
\end{array}
\right)}
\def\twovec#1#2{\left(\begin{array}{c}
{#1}\\ {#2}\\
\end{array}
\right)}
\begin{titlepage}
\hskip 5.5cm
\hskip 1.5cm \vbox{\hbox{CERN-TH/2000-167}\hbox{hep-th/0006xx}\hbox{June, 2000}} \vfill
\vskip 3cm
\begin{center}
{\LARGE {  Matter Coupled $F(4)$ Supergravity and the $AdS_6/CFT_5$ Correspondence}}\\
\vskip 1.5cm
  {\bf Riccardo D'Auria$^1$, Sergio Ferrara$^2$ and
Silvia Vaul\'a$^3$ } \\
\vskip 0.5cm {\small $^1$ Dipartimento di Fisica, Politecnico di
Torino,\\
 Corso Duca degli Abruzzi 24, I-10129 Torino\\
and Istituto Nazionale di Fisica Nucleare (INFN) - Sezione di
Torino, Italy}\\
{\small $^2$ CERN Theoretical Division, CH 1211 Geneva 23, Switzerland }\\
{\small $^3$ Dipartimento di Fisica Teorica, Universit\'a di
Torino, via P. Giuria 1, I-10125 Torino} \vspace{6pt}
\end{center}
\vskip 3cm \vfill
\begin{center} {\bf Abstract}
\end{center}
{\small
$F(4)$  supergravity, the gauge theory of the exceptional
six-dimensional Anti-de Sitter superalgebra, is coupled  to an
arbitrary number of vector multiplets whose scalar components
parametrize the quaternionic manifold $SO(4,n)/SO(4)\times SO(n)$. By
gauging the compact subgroup $SU(2)_d \otimes \cG$, where
$SU(2)_d$ is the diagonal subgroup of $SO(4)\simeq SU(2)_L\otimes
SU(2)_R$ (the$R$-symmetry group of six-dimensional Poincar\'e
supergravity) and $\cG$ is a compact group such that $dim\cG = n$,
 we compute the scalar potential which, besides the gauge
coupling constants, also depends in non trivial way on the
parameter $m$ associated to a massive 2-form $B_{\mu\nu}$ of the
gravitational multiplet. The potential admits an $AdS$ background
for $g=3m$, as the pure $F(4)$-supergravity. We compute the
scalar squared masses (which are all negative) and  retrieve the
results dictated by $AdS_6/CFT_5$ correspondence from the
conformal dimensions of boundary operators. The boundary $F(4)$
superconformal fields are realized in terms of a singleton
superfield (hypermultiplet) in harmonic superspace with flag
manifold $SU(2)/U(1)=S^2$. We analize the spectrum of short
representations in terms of  superconformal primaries and
predict general features of the K-K specrum of massive type
IIA supergravity compactified on warped $AdS_6\otimes S^4$.}

\end{titlepage}
\section{Introduction}

In the past two years much work  has been devoted to the
verification of the conjectured duality between $AdS_{d+1}$
supergravities and conformal field theories in Minkowski
$d-$dimensional space \cite{malda} (for an extensive review of
the topic see \cite{rass}). In this framework an important check
is given by comparing the Kaluza-Klein mass spectrum of
compactified $M$ and/or superstring theories in lower dimensions,
with the mass spectrum computed in terms of the scale dimension
$\Delta\equiv E_0$  of the conformal operators of the dual
theory. While much effort has been devoted to the investigation
of such correspondence for seven, five and four dimensional
dimensional supergravities, the six dimensional case based on the
$F(4)$ exceptional supergroup \cite{nahm}, \cite{bagu} has been studied only in
reference \cite{fkpz}. In this paper it was shown that the $F(4)$
theory has on its boundary a superconformal field theory at a
fixed point of the five-dimensional Yang-Mills theory. The relevant
superconformal operators in five dimensions, which are dual to the
supergravity states in $D=6$, are built in terms of the singleton
representation of $F(4)\otimes\mathcal{G}$ that is just a
5-dimensional hypermultiplet carring a representation
of the flavour group $\mathcal{G}$.\\
In this way, it is possible to predict the mass spectrum of the
"massless" supergravity states in a supersymmetric $AdS$
background from the corresponding $E_0$ quantum number of the
conformal supermultiplet. This prediction, however, could not be
checked since the matter coupled $F(4)$ theory was not yet constructed;
 what one usually refers to as $F(4)$ supergravity is the
theory constructed  by Romans \cite{rom} in the pure supergravity
case, that is, without matter coupling. (Some aspects of the
$F(4)$ theory have also been discussed in the framework of
dimensional reduction from seven dimensions in \cite{vn}).\\
Other interesting results on the $F(4)$ theory were obtained in
reference \cite{oz} where it was shown that the theory obtained
from spontaneous compactification of Massive Type IIA
Supergravity in ten dimensions down to the $F(4)$ gauged
supergravity in six dimensions can be described in terms of the
near horizon geometry of the $D4-D8$ brane system. Furthermore in
reference \cite{clp} it was shown that the $F(4)$ gauged pure
supergravity constructed in ref \cite{rom}, can be obtained as
a consistent warped  reduction on $S^4$ from massive Type IIA ten
dimensional Supergravity \cite{mtIIA}.\\
The $AdS_6/CFT_5$ correspondence further predicts that the $K-K$
excitations of massive type IIA on $AdS_6\otimes S^4$ should be
related to towers of superconformal primary  operators of the
boundary superconformal field theory. The latter having only
eight Poincar\'e supercharges is not completely fixed by
supersymmetry and in fact it can have  various global symmetry
groups $\cG$ which were classified in \cite{smi}. This reflects in the gauge groups of the vector
multiplets of $F(4)$ whose coupling is discussed here.\\
It will be shown that the massless graviton (stress-energy
tensor) multiplet and the "current multiplet" related to the
$\cG$ gauge fields, previously discussed in \cite{fkpz}, are
actually the first members of two distinct towers of (short)
conformal superfields, one corresponding to the $1/2$ BPS
multiplets (massive vector multiplets), the other corresponding
to the graviton recurrences.\\
Interestingly enough they correspond to two isolated classes of
highest-weight UIR's of the $F(4)$ superalgebra, separated from
the continuous spectrum (in $E_0$) suggested to exist in
\cite{min}.\\
\newpage
The paper is organized as follows:\\

In section 2 the geometric approach, and the closely related
superspace solutions of Bianchi identities are presented in the
framework of the pure $F(4)$ supergravity of ref. \cite{rom}.
This, besides to give the right geometrical setting for the matter coupling
and gauging of the theory, allows us to discuss in a simple way
the algebraic foundation of the peculiar properties of $F(4)$
superalgebra namely, the  relations between the $SU(2)$ gauging coupling constant
$g$ and the inverse $AdS$ radius $4m$ and the related "Higgs
phenomenon" by which the gravitational two-form $B_{\mu\nu}$
becomes massive.\\
\indent In section 3 the geometrical framework developed in section 2 is
applied to the matter coupling and in section 4 the gauging of $D=6$ $N=(1,1)$ supergravity is studied.\\
\indent In section 5 the scalar potential is derived and it is shown to admit an $AdS_6$ supersymmetric vacuum.\\
\indent In section 6 the boundary conformal field
theory and its short representations in terms of harmonic
superfields \cite{fanta1} (with harmonic space $S^2=SU(2)/U(1)$) are presented
and the spectrum of K-K excitations predicted.

\section{The geometrical approach}
In this section we set up a suitable framework for the discussion
of the matter coupled $F(4)$ supergravity theory and its gauging.
\noindent This will allow us to set up the formalism for the
matter coupling in the next section. Actually we will just give
the essential definitions of the Bianchi identities approach in
superspace , while all the relevant results, specifically the
supersymmetry transformation laws of the fields, will be given in
the ordinary space-time formalism.

First of all it is useful to discuss the main results of ref.  \cite{rom} by a careful study in superspace of the
Poincar\'e and AdS supersymmetric vacua.
 Let us recall the content of $D=6$, $N=(1,1)$ supergravity
multiplet:
\begin{equation}
(V^{a}_{\mu}, A^{\alpha}_{\mu}, B_{\mu\nu},\,\ \psi^{A}_{\mu},\,\
\psi^{\dot{A}}_{\mu}, \chi^{A}, \chi^{\dot{A}}, e^{\sigma})
\end{equation}
\noindent where $V^a_{\mu}$ is the six dimensional vielbein,
$\psi^{A}_{\mu},\,\ \psi^{\dot{A}}_{\mu}$ are left-handed and
right-handed four- component gravitino fields respectively, $A$
and $\dot{A}$ transforming under the two factors of the
$R$-symmetry group $O(4)\simeq SU(2)_L\otimes SU(2)_R$,
$B_{\mu\nu}$ is a 2-form, $A^{\alpha}_{\mu}$ ($\alpha=0,1,2,3$),
are  vector fields, $\chi^{A}, \chi^{\dot{A}}$ are left-handed and
right-handed spin $\frac{1}{2}$ four components dilatinos, and
$e^{\sigma}$ denotes the dilaton.\\ Our notations are as follows:
$a,b,\dots=0,1,2,3,4,5$ are Lorentz flat indices in $D=6$
$\mu,\nu,\dots=0,1,2,3,4,5$ are the corresponding world indices,
  $A,\dot{A}=1,2$. Moreover our metric is
$(+,-,-,-,-,-)$.\\ We recall that the description of the spinors
of the multiplet in terms of left-handed and right-handed
projection holds only in a Poincar\'e background, while in an AdS
background the chiral projection cannot be defined and we are
bounded to use 8-dimensional pseudo-Majorana spinors. In this
case the $R$-symmetry group reduces to the $SU(2)$ subgroup of
$SU(2)_L\otimes SU(2)_R$, the $R$-symmetry group of the chiral
spinors. For our purposes, it is convenient to use from the very
beginning 8-dimensional pseudo-Majorana spinors even in a
Poincar\'e framework, since we are going to discuss in a unique
setting both Poincar\'e and $AdS$ vacua.\\ The pseudo-Majorana
condition on the gravitino 1-forms is as follows:
\begin{equation}
(\psi_A)^{\dagger}\gamma^0=\overline{(\psi_A)}=\epsilon^{AB}\psi_{B}^{\
\ t}
\end{equation}
\noindent where we have chosen the charge conjugation matrix in
six dimensions as the identity matrix (an analogous definition
six dimensions as the identity matrix (an analogous definition
holds for the dilatino fields). We use eight dimensional
antisymmetric gamma matrices, with
$\gamma^7=i\gamma^0\gamma^1\gamma^2\gamma^3\gamma^4\gamma^5$,
which implies $\gamma_7^T=-\gamma_7$ and $(\gamma_7)^2=-1$. The
indices $A,B,\dots=1,2,$ of the spinor fields $\psi_A,\,\ \chi_A$
transform in the fundamental of the diagonal subgroup $SU(2)$ of
$SU(2)_L\otimes SU(2)_R$. For a generic $SU(2)$ tensor $T$,
raising and lowering of indices are defined by
\begin{eqnarray}
&&T^{\dots A\dots}=\epsilon^{AB}\ \ T^{\dots\ \ \dots}_{\ \ B}\\
&&T_{\dots A\dots}=T_{\dots\ \ \dots}^{\ \ B}\ \ \epsilon_{BA}
\end{eqnarray}

 To study the supersymmetric vacua let us write down the Maurer-Cartan
Equations (M.C.E.) dual to the $F(4)$ Superalgebra
(anti)commutators:

\begin{eqnarray}
\label{dVx}&&\mathcal{D}V^{a}-\frac{i}{2}\ \
\overline{\psi}_{A}\gamma_{a}\psi^A =0\\
&&\mathcal{R}^{ab}+4m^{2}\ \
V^{a}V^{b}+m\overline{\psi}_{A}\gamma_{ab}\psi^A=0\\
&&dA^{r}+\frac{1}{2}\,\ g\,\ \epsilon^{rst}A_{s}A_{t}-i\,\
\overline{\psi}_{A}\psi_{B}\,\
\sigma^{rAB}=0\\
\label{dPsix}&&D\psi_{A}-im\gamma_{a}\psi_{A}V^{a}=0
\end{eqnarray}

\noindent Here $V^a,\omega^{ab},\psi_A,A^r,(r= 1,2,3)$, are
superfield 1-forms dual to the $F(4)$ supergenerators  which at
$\theta =0$ have as $dx^{\mu}$ components
\begin{equation}
V^a_{\mu}= \delta^a_{\mu},\,\ \psi_{A\mu}=A^r_{\mu}=0, \,\
\omega^{ab}_{\mu}= pure\,\ gauge.
\end{equation}
Furthermore $\mathcal{R}^{ab}\equiv
d\omega^{ab}-\omega^{ac}\land\,\ \omega_c^{\,\ b}$, $\mathcal{D}$
is the Lorentz covariant derivative, $D$ is the $SO(1,5)\otimes
SU(2)$ covariant derivative, which on spinors acts as follows:
\begin{equation}D\psi_A\equiv
d\psi_A-\frac{1}{4}\gamma_{ab}\omega^{ab}\psi_A-\frac{i}{2}\sigma_{AB}^r
A_r\psi^B
\end{equation}
\noindent Note that $\sigma^{rAB}=\epsilon^{BC}\sigma^{rA}_{\ \
C}$, where $\sigma^{rA}_{\ \ B}$\ \ ($r=1,2,3$) denote the usual
Pauli matrices, are symmetric in $A,\,\ B$.\\ Let us point out
that the $F(4)$ superalgebra, despite the presence of two
different physical parameters, the $SU(2)$ gauge coupling constant
$g$ and the inverse $AdS$ radius $m$, really depends on just one
parameter since the closure under $d$-differentiation of eq.
(\ref{dPsix})  (equivalent to the implementation of Jacobi
identities on the generators), implies $g=3m$; to recover this
result one has to use the following Fierz identity involving
3-$\psi_A$'s 1-forms:
\begin{equation}
\label{fond}\frac{1}{4}\gamma_{ab}\psi_A\overline{\psi}_B\gamma^{ab}\psi_C\epsilon^{AC}-\frac{1}{2}
\gamma_{a}\psi_A\overline{\psi}_B\gamma^{a}\psi_C\epsilon^{AC}+3\psi_C\overline{\psi}_B\psi_A\epsilon^{BC}=0
\end{equation}\\The $F(4)$ superalgebra described by equations
(\ref{dVx}) -(\ref{dPsix}) fails to describe the physical vacuum
because of the absence of the superfields 2-form $B$ and 1-form
$A^0$ whose space-time restriction coincides with the physical
fields $B_{\mu\nu}$ and
$A_{\mu}^0$ appearing in the supergravity multiplet.
 The recipe to have all the fields in
a single algebra is well known and consists in considering the
Free Differential Algebra (F.D.A.)\cite{bible} obtained from the
$F(4)$ M.C.E.'s by adding two more equations for the 2-form $B$
and for the 1-form $A^0$ (the 0-form fields $\chi_A$ and $\sigma$
do not appear in the algebra since they are set equal to zero in
the vacuum). It turns out that to have a consistent F.D.A.
involving $B$ and $A^0$ one has to add to the $F(4)$ M.C.E.'s two
more equations involving $dA^0$ and $dB$; in this way one obtains
an extension of the M.C.E's to the following F.D.A:
\begin{eqnarray}\label{dV}&&\mathcal{D}V^{a}-\frac{i}{2}\ \
\overline{\psi}_{A}\gamma_{a}\psi^A=0 \\
\label{dO}&&\mathcal{R}^{ab}+4m^{2}\ \
V^{a}V^{b}+m\overline{\psi}_{A}\gamma_{ab}\psi^A=0\\
 \label{dAr}&&dA^{r}+\frac{1}{2}\,\ g\,\
\epsilon^{rst}A_{s}A_{t}-i\,\ \overline{\psi}_{A}\psi_{B}\,\
\sigma^{rAB}=0\\
\label{dA}&&dA^0-mB-i\,\
\overline{\psi}_{A}\gamma_{7}\psi^A=0\\
\label{dB}&&dB+2\,\
\overline{\psi}_{A}\gamma_{7}\gamma_{a}\psi^AV^{a}=0\\
\label{dPsi}&&D\psi_{A}-im\gamma_{a}\psi_{A}V^{a}=0\end{eqnarray}
\noindent  Equations (\ref{dA}) and (\ref{dB}) were obtained by
imposing that they satisfy the $d$-closure together with equations
(\ref{dV}). Actually the closure of (\ref{dB}) relies on the
4-$\psi_A$'s Fierz identity
\begin{equation}
\overline{\psi}_{A}\gamma_{7}\gamma_{a}\psi_{B}\epsilon^{AB}\overline{\psi}_{C}\gamma^{a}\psi_{D}\epsilon^{CD}=0
\end{equation}
The interesting feature of the F.D.A (\ref{dV})-(\ref{dPsi}) is
the appearance of the combination $dA^0-mB$ in (\ref{dA}). That
means that the dynamical theory obtained by gauging the F.D.A.
out of the vacuum will contain the fields $A^0_{\mu}$ and
$B_{\mu\nu}$ always in the single combination
$\partial_{[\mu}A^0_{\nu]}-mB_{\mu\nu}$. At the dynamical level
this implies, as noted by Romans \cite{rom}, an Higgs phenomenon
where the 2-form $B$ "eats" the 1-form $A^0$ and acquires a non
vanishing mass $m$ \footnote{An analogous phenomenon takes place
also in $D=5$; see\cite{anna}.}.\\ In summary, we have shown that
two of the main results of \cite{rom}, namely  the existence of
an $AdS$ supersymmetric background only for $g=3m$ and  the
Higgs-type mechanism by which the field $B_{\mu\nu}$ becomes
massive acquiring longitudinal degrees of freedom in terms of the
the vector $A^0_{\mu}$, are a simple consequence of the algebraic
structure of the F.D.A. associated to the $F(4)$ supergroup
written in terms of the
vacuum-superfields.\\
It is interesting to see what happens if one or both the
parameters $g$ and $m$ are zero. Setting $m=g=0$, one reduces the
$F(4)$ Superalgebra to the $D=6\ \ N=(1,1)$ superalgebra existing
only in a Super Poincar\'e background; in this case the four-
vector $A^{\alpha}\equiv (A^0,A^r)$ transforms in the fundamental
of the $R$-symmetry group $SO(4)$ while the pseudo-Majorana spinors
$\psi_A,\chi_A$ can be decomposed in two chiral spinors in such a
way that all the resulting F.D.A. is invariant under
$SO(4)$.\\
Furthermore it is easy to see that no F.D.A  exists if either
$m=0$ , $g\neq 0$ or $m\neq 0$, $g= 0$, since the corresponding
equations in the F.D.A. do not close anymore under $d$-
differentiation. In other words the gauging of $SU(2)$, $g\neq 0$
must be necessarily accompanied by the presence of the parameter
$m$ which, as we have
 seen, makes the closure of the supersymmetric algebra consistent for $g=3m$.\\

Let us now consider the dynamical theory out of the vacuum. For
the sake of completeness and in order to establish a consistent
notation, we first reconsider the pure supergravity theory of
ref. \cite{rom} in a superspace formalism. We start from the
F.D.A. of $D=6,\ \ N=(1,1)$, $SU(2)$-gauged Poincar\'e
Supergravity, namely
we consider the  F.D.A. (\ref{dV}) -(\ref{dPsi}) taking $m=0$, $g\neq 0$.\\
Following the geometrical approach \cite{bible}, we define the
supercurvatures as deviation from the $m=0$, $g\neq 0$ F.D.A. as
follows: {\setlength\arraycolsep{1pt}\begin{eqnarray}
\label{T}&T^a&=\mathcal{D}V^a-\frac{i}{2}\ \
\overline{\psi}_A\gamma_a\psi^AV^a=0\\
\label{L}&R^{ab}&=\mathcal{R}^{ab}\\ \label{H} &H&=dB+2
e^{-2\sigma}\,\
\overline{\psi}_A\gamma_7\gamma_a\psi^AV^a\\
 \label{F}&F&=dA-ie^{\sigma}\,\
\overline{\psi}_A\gamma_7\psi^A\\
\label{Fr}&F^r&=dA^r+\frac{1}{2}\,\ g\,\
\epsilon^{rst}A_sA_t-ie^{\sigma}\,\ \overline{\psi}_A\psi_B\,\
\sigma^{rAB}\\ \label{ro} &\rho_A&=D\psi_A\\
 \label{R}&R(\chi_A)&\equiv D\chi_A\\
 \label{s}&R(\sigma)&\equiv
 d\sigma
\end{eqnarray}}

\noindent where we have implemented the kinematical constraint
that the supertorsion $T^a$ is identically zero, and we have added
the "curvatures" of the $0$-forms $\chi_A$ and $\sigma$ defined as
their (covariant) differential.\\ Differentiating the curvatures
one obtains the Bianchi identities

\begin{eqnarray}
\label{bT}&&R^{ab} V_b-i
\overline{\psi}_A\gamma^a\rho^A=0\\
\label{bL}&&\mathcal{D}R^{ab}=0\\
\label{bH}&&dH\!+\!4e^{-2\sigma}d\sigma\,\
\overline{\psi}_A\gamma_7\gamma_a\psi^AV^a\!+\!4e^{-2\sigma}
\overline{\psi}_A\gamma_7\gamma_a\rho^AV^a\!=\!0\\
 \label{bF}&&DF+id\sigma e^{\sigma}\,\
\overline{\psi}_A\gamma_7\psi^A-2ie^{\sigma}\,\
\overline{\psi}_A\gamma_7\rho^A=0\\
\label{bFr}&&DF^r+id\sigma e^{\sigma}\,\
\overline{\psi}_A\psi_B\,\ \sigma^{rAB}-2ie^{\sigma}\,\
\overline{\psi}_A\rho_B\,\ \sigma^{rAB}=0\\
\label{bro}&&D^2\psi_A+\frac{1}{4}R^{ab}\gamma_{ab}\psi_A-\frac{i}{2}\,\
g\,\ \sigma_{rAB}F^r\psi^B=0\\
\label{bR}&&D^2\chi_A+\frac{1}{4}R^{ab}\gamma_{ab}\chi_A-\frac{i}{2}\,\
g\,\ \sigma_{rAB}F^r\chi^B=0\\ \label{bs}&&d^2\sigma=0
\end{eqnarray}

The superspace solutions of the previous Bianchi Identities are
given by:

{\setlength\arraycolsep{1pt}\begin{eqnarray}
\label{cip}&R^{ab}&=\widetilde{R}^{ab}_{cd}
V^cV^d+\overline{\theta}^{ab}_{cA}\psi^A V^c+\overline{\psi}_A
M^{ab}\psi^A+\frac{1}{4}g\,\
e^{\sigma}\overline{\psi}_A\gamma^{ab}\psi^A\\
\label{minni}&H&=\widetilde{H}_{abc}V^aV^bV^c+4i\,\ e^{-2\sigma}\
\ \overline{\psi}_{A}\gamma_{7}\gamma_{ab}\chi^A\ \
 V^aV^b\\
 &F&=\widetilde{F}_{ab}V^{a}V^{b}+2\,\ e^{\sigma}\ \
 \overline{\psi}_{A}\gamma_{7}\gamma_{a}\chi^AV^{a}\\
 &F^{r}&=\widetilde{F}^{r}_{ab}\,\ V^{a}V^{b}+2\,\ e^{\sigma}\ \
 \overline{\psi}_{A}\gamma_{a}\chi_{B}\ \
\sigma^{rAB}\,\ V^{a}\\  \label{obiwan}& D \psi_A &=\widetilde{
D_{[a} \psi_{b]A}} V^{a}V^{b}+\frac{1}{16}
e^{-\sigma}[\epsilon_{AB}\,\ \widetilde{F}_{ab}\gamma_{7}-
\sigma_{rAB}\,\
\widetilde{F}^{r}_{ab}](\gamma^{cab}-6\delta^{ca}\gamma^{b})\,\
\psi^{B} V_{c}+\nonumber\\ &&+\frac{i}{32} e^{2\sigma}
\widetilde{H}_{abc}\,\ \gamma_{7}(\gamma^{dabc}-3\delta^{ad}
\gamma^{bc})\,\ \psi_{A} V_{d}-\frac{i}{4}g\,\ e^{\sigma}
\gamma_{a}\psi_{A}
V^{a}\nonumber\\
&&+\frac{1}{4}\gamma_{7}\psi_{A}\overline{\chi}^{C}\gamma^{7}\psi_{C}-\frac{1}{2}\gamma_{a}\psi_{A}\overline{\chi}^{C}\gamma^{a}\psi_{C}+\frac{1}{2}\gamma_{7}\gamma_{a}\psi_{A}\overline{\chi}^{C}\gamma^{7}\gamma^{a}\psi_{C}+\nonumber\\
&&-\frac{1}{8}\gamma_{ab}\psi_{A}\overline{\chi}^{C}\gamma^{ab}\psi_{C}-\frac{1}{8}\gamma_{7}\gamma_{ab}\psi_{A}\overline{\chi}^{C}\gamma^{7}\gamma^{ab}\psi_{C}+\frac{1}{4}\psi_{A}\overline{\chi}^{C}\psi_{C}\\
\label{mfw}&D\chi_{A}&=\widetilde{D_{a} \chi_{A}} V^{a}+
\frac{i}{2} \gamma^{a}\widetilde{\partial_{a}\sigma} \,\ \psi_{A}+
\frac{i}{16} e^{\sigma}[\epsilon_{AB} \,\
\widetilde{F}_{ab}\gamma_{7}+\sigma_{rAB} \,\
\widetilde{F}^{r}_{ab}]\gamma^{ab}\psi^{B}\nonumber\\
&&+\frac{1}{32} e^{2\sigma}
\widetilde{H}_{abc}\gamma_{7}\gamma^{abc}\psi_{A}-\frac{1}{4}g
e^{\sigma}\psi_{A}\\
\label{ciop}&d\sigma&=\widetilde{\partial_{a}\sigma}
V^{a}+\overline{\chi}_{A}\psi^A
\end{eqnarray}}

\noindent where we have no written down the explicit form of
$\overline{\theta}^{ab}_{cA}$ and $M^{ab}$ since they are rather
cumbersome and uninteresting for our later purposes.\footnote
{Note that the tilded quantities correspond, on space-time, to the
supercovariant field strengths: namely e.g. projecting equation
(\ref{minni}) along the space time differentials $dx^{\mu}$'s we
have: $\widetilde{H}_{\mu\nu\rho}=H_{\mu\nu\rho}-4i\,\
e^{-2\sigma}\ \
\overline{\psi}_{A[\mu}\gamma_{\nu\rho]}\gamma_{7}\chi^A$ that
is, using definition (\ref{H})
$\widetilde{H}_{\mu\nu\rho}=\partial_{[\mu}B_{\nu\rho]}+2
e^{-2\sigma}\,\
\overline{\psi}_{A[\mu}\gamma^7\gamma_{\rho}\psi^A_{\nu]}-4i\,\
e^{-2\sigma}\ \
\overline{\psi}_{A[\mu}\gamma_{\nu\rho]}\gamma_{7}\chi^A$.
Analogous formulae hold for the other tilded quantities.} Note
that setting $g=0$ in the fermion fields solutions the
corresponding superspace curvatures give the solutions of the
Bianchi identities of $N=(1,1)$ $D=6$ Poincar\'e Supergravity (in
this case of course the covariant derivatives in the l.h.s. are
Lorentz covariant and the $\mathcal{F}^r$ is abelian).
\\At $g\neq 0$, however, one can immediately
see that the vacuum configuration corresponding to this solution
is not supersymmetric. Indeed if we set
\begin{equation}
\label{vacca}
\widetilde{F}_{ab}=\widetilde{F}^r_{ab}=\widetilde{H}_{abc}=\sigma=\chi_A=
\psi_{A\mu}=0
\end{equation}
\noindent we see that the superspace field strengths $D\psi_A$,
$D\chi_A$ are not zero in the vacuum if $g\neq 0$, due to the
presence of the gauging terms proportional to $g$ in their
superspace parametrisations. In the ordinary space-time language
that means that, in the vacuum, the supersymmetry transformation
law of the dilatino is not zero and that the gravitino doesn't
transforms as the Lorentz covariant derivative of the
supersymmetry parameter $\varepsilon_A$:
\begin{eqnarray}
&&\delta\chi_A=-\frac{1}{4}g e^{\sigma}\varepsilon_{A}\\
&&\delta\psi_{A\mu}=\mathcal{D}_{\mu}\varepsilon_A-\frac{i}{4}g\,\
e^{\sigma} \gamma_{\mu}\varepsilon_{A}
\end{eqnarray}
This is of course in line with what we have found previously in
the study of the supersymmetric vacua, that is the fact that at
$g=0$ we have suitable supersymmetric Poincar\'e vacuum, while we
are bound to modify the previous solution if we want to obtain a
supersymmetric $AdS$ vacuum.\\
This modification however is very simple to obtain since we
already know that the proper vacuum configuration of the $F(4)$
theory requires a F.D.A modified by the mass parameter $m$
suitable related to $g$, as we have shown in the previous
discussion.\\
Therefore to obtain such vacuum we modify the r.h.s. of equations
(\ref{cip}) -(\ref{ciop}) by adding suitable $m$ terms in such a
way that the Bianchi identities (\ref{bT}) -(\ref{bs}) are still
satisfied. To study the supersymmetric vacua of the new solutions,
it is convenient to denote with the suffix $(g=0)$ the fermionic
field strengths and the Lorentz curvature in equations (\ref{cip})
-(\ref{ciop}) evaluated at $g=0$. The new solution extending the
previous one with terms containing $m$ is given by:
\begin{eqnarray}
&&R^{ab\,\ (new)}=R^{ab\,\ (g=0)}+4m^2\,\
e^{-6\sigma}V^aV^b+\frac{1}{4}m\,\
e^{-3\sigma}\overline{\psi}_A\gamma^{ab}\psi^A+\nonumber\\
&& +\frac{1}{4}g\,\ e^{\sigma}\overline{\psi}_A\gamma^{ab}\psi^A\\
&&H^{(new)}=H\\
&&F^{(new)}=F-mB\\
&&F^{r\,\ (new)}=F^{r}\\
&&\rho_A^{(new)}=\rho_A^{(g=0)}-\frac{i}{4}
me^{-3\sigma}\gamma_a\psi_A V^a -\frac{i}{4}
ge^{\sigma}\gamma_a\psi_A V^a \\
&&D\chi_A^{(new)}=D\chi_A^{(g=0)}+\frac{3}{4}
me^{-3\sigma}\psi_A-\frac{1}{4}
ge^{\sigma}\psi_A\\
&&d\sigma^{(new)}=d\sigma
\end{eqnarray}
\noindent It is now immediate to see that, in the vacuum defined
by equation (\ref{vacca}), the dilatino field strength vanishes
only for $g=3m$. Furthermore in this case, the extra $g$ and $m$
terms in the gravitino field strength are the correct ones in
order to reconstruct  the vanishing of the $AdS$ covariant
gravitino curvature while the extra term $-mB$ in $F^{(new)}$ and
$4m^2\,\ V^aV^b+m\overline{\psi}_A\gamma^{ab}\psi^A$ in $R^{ab\,\
(new)}$ are exactly what is needed in order to have a
supersymmetric $AdS$ background (see eqs. (\ref{dV})
-(\ref{dPsi})).\\ From the superspace solution of the Bianchi
identities one immediately derives the space-time supersymmetry
transformation law of the physical fields:
{\setlength\arraycolsep{1pt}\begin{eqnarray}&\delta
V^{a}_{\mu}&=-i\overline{\psi}^A_{\mu}\gamma^{a}\varepsilon^A\\
&\delta B_{\mu\nu}&=4i\,\
e^{-2\sigma}\overline{\chi}_{A}\gamma_{7}\gamma_{\mu\nu}\varepsilon^A-4e^{-2\sigma}\overline{\varepsilon}_A\gamma_7\gamma_{[\mu}\psi_{\nu]}^A\\
 &\delta A_{\mu}&=- 2\,\ e^{\sigma}
 \overline{\chi}_{A}\gamma_{7}\gamma_{\mu}\varepsilon^A
 +2ie^{\sigma}\overline{\varepsilon}_A\gamma^7\psi^A_{\mu}\\
 &\delta A^{r}_{\mu}&=2e^{\sigma}
 \overline{\chi}^{A}\gamma_{\mu}\varepsilon^{B}
\sigma^{r}_{AB}+2ie^{\sigma}\sigma^{rAB}\overline{\varepsilon}_A\psi_{B\mu}\\
\label{bart}&\delta\psi_{A\mu}&=D_{\mu}\varepsilon_{A}+\frac{1}{16}
e^{-\sigma}[\epsilon_{AB} \widetilde{F}_{\nu\lambda}\gamma_{7}-
\sigma_{rAB}
\widetilde{F}^{r}_{\nu\lambda}](\gamma_{\mu}^{\nu\lambda}-6\delta_{\mu}^{\nu}\gamma^{\lambda})\,\
\varepsilon^{B}+\nonumber\\
&&+\frac{i}{32} e^{2\sigma} \widetilde{H}_{\nu\lambda\sigma}\,\
\gamma_{7}(\gamma_{\mu}^{\,\ \nu\lambda\sigma}-3\delta_{\mu}^{\nu}
\gamma^{\lambda\sigma}) \varepsilon_{A}-\frac{i}{4}g\,\
e^{\sigma} \gamma_{\mu}\varepsilon_{A} -\frac{i}{4}
me^{-3\sigma}\gamma_{\mu}\varepsilon_A\nonumber\\
&&+\frac{1}{2}\gamma_{7}\varepsilon_{A}\overline{\chi}^{C}\gamma^{7}\psi_{\mu
C}-\gamma_{\nu}\varepsilon_{A}\overline{\chi}^{C}\gamma^{\nu}\psi_{\mu
C}+\gamma_{7}\gamma_{\nu}\varepsilon_{A}\overline{\chi}^{C}\gamma^{7}\gamma^{\nu}\psi_{\mu
C}+\nonumber\\
&&-\frac{1}{4}\gamma_{\nu\lambda}\varepsilon_{A}\overline{\chi}^{C}\gamma^{\nu\lambda}\psi_{\mu
C}-\frac{1}{4}\gamma_{7}\gamma_{\nu\lambda}\varepsilon_{A}\overline{\chi}^{C}\gamma^{7}\gamma^{\nu\lambda}\psi_{\mu
C}+\frac{1}{2}\varepsilon_{A}\overline{\chi}^{C}\psi_{\mu C}\\
\label{homer}&\delta\chi_{A}&= \frac{i}{2}
\gamma^{\mu}\widetilde{\partial_{\mu}\sigma}\varepsilon_{A}+
\frac{i}{16} e^{\sigma}[\epsilon_{AB}
\widetilde{F}_{\mu\nu}\gamma_{7}+\sigma_{rAB}
\widetilde{F}^{r}_{\mu\nu}]\gamma^{\mu\nu}\varepsilon^{B}\nonumber\\
&&+\frac{1}{32} e^{2\sigma}
\widetilde{H}_{\mu\nu\lambda}\gamma_{7}\gamma^{\mu\nu\lambda}\varepsilon_{A}-\frac{1}{4}g
e^{\sigma}\varepsilon_{A}+\frac{3}{4} me^{-3\sigma}\varepsilon_A\\
&\delta\sigma&=\overline{\chi}_{A}\varepsilon^A
\end{eqnarray}
Apart from different conventions and normalizations the above
equations coincide with the results of reference \cite{rom},
except for the extra terms in the gravitino transformation law of
the form $\psi \chi \varepsilon$, which, like all the
three-fermion terms, were not computed in reference \cite{rom}.
However these terms correspond to terms $\psi \psi \chi$ in the
superspace curvature $D\psi_A$ which are quite essential to verify
the consistency of the Bianchi identities when the parameter $m$
is introduced and a supersymmetric $AdS$ background is found;
therefore they have an important meaning for the consistence of
the theory and this is the reason why we have explicitly quoted
them. This is to be contrasted with other three-fermion terms of
the form $\chi \chi \varepsilon$ on space-time ($\chi \chi \psi$
in superspace), which we have not included in the transformation
law of the fermions, since their explicit form can be found from
the Bianchi identities once the consistency in the higher sectors
has been verified, so that they are not on the same status.
\noindent In the supersymmetric $AdS$ vacuum we get:
\begin{eqnarray}
&&\delta\chi_A=0\\
&&\delta\psi_{A\mu}=\nabla^{AdS}_{\mu}\epsilon_A\\
\label{curv}&&R^{ab}\equiv -\frac{1}{2} R^{ab}_{cd}V^cV^d=-4m^2 V^aV^b\rightarrow R_{\mu\nu}=20m^2g_{\mu\nu}\\
&&\mathcal{F}^r_{\mu\nu}=\mathcal{F}_{\mu\nu}-mB_{\mu\nu}=\chi_A=\psi_{A\mu}=\sigma=0
\end{eqnarray}

\section{Coupling to matter multiplets}
In $D=6,\,\ N=4$ Supergravity, the only kind of matter is given
by vector multiplets, namely
\begin{equation}
(A_{\mu},\,\ \lambda_A,\,\ \phi^{\alpha})^I
\end{equation}
\noindent where $\alpha=0,1,2,3$ and the index $I$ labels an
arbitrary number $n$ of such multiplets. As it is well known the
$4n$ scalars parametrize the coset manifold $SO(4,n)/SO(4)\times
SO(n)$. Taking into account that the pure supergravity has a non
compact duality group $O(1,1)$ parametrized by $e^{\sigma}$, the
duality group of the matter coupled theory is
 \begin{equation}\label{coset}
  G/H=\frac{SO(4,n)}{SO(4)\times SO(n)}\times O(1,1)
\end{equation}
To perform the matter coupling we follow the geometrical
procedure of introducing the coset representative $L^{\Lambda}_{\
\ \Sigma}$ of the matter coset manifold, where
$\Lambda,\Sigma,\dots=0, \dots, 3+n$; decomposing the $O(4,n)$
indices with respect to $H=SO(4)\times O(n)$ we have:
\begin{equation}
L^{\Lambda}_{\ \ \Sigma}=(L^{\Lambda}_{\ \ \alpha},L^{\Lambda}_{\
\ I})
\end{equation}
\noindent where $\alpha=0,1,2,3$, $I=4,\dots ,3+n$. Furthermore,
since we are going to gauge the $SU(2)$ diagonal subgroup of
$O(4)$ as in pure Supergravity, we will also decompose
$L^{\Lambda}_{\ \ \alpha}$ as
\begin{equation}
L^{\Lambda}_{\ \ \alpha}=(L^{\Lambda}_{\ \ 0}, L^{\Lambda}_{\ \
r})
\end{equation}
The $4+n$ gravitational and matter vectors will now transform in
the fundamental of $SO(4,n)$ so that the superspace vector
curvatures will be now labeled by the index $\Lambda$:
$F^{\Lambda} \equiv (F^0,F^r,F^I)$. Furthermore the covariant
derivatives acting on the spinor fields will now contain also the
composite connections of the $SO(4,n)$ duality group. Let us
introduce the left-invariant 1-form of $SO(4,n)$
\begin{equation}
\Omega^{\Lambda}_{\ \
\Sigma}=(L^{\Lambda}_{\ \ \Pi})^{-1} dL^{\Pi}_{\ \
\Sigma}
\end{equation}
 \noindent satisfying the Maurer-Cartan
equation
\begin{equation}
d\Omega^{\Lambda}_{\ \ \Sigma}+\Omega^{\Lambda}_{\ \
\Pi}\land\Omega^{\Pi}_{\ \ \Sigma}=0
\end{equation}
\noindent By
appropriate decomposition of the indices, we find:
\begin{eqnarray}
\label{1}&&R^r_{\,\ s}=-P^{r}_{\ \ I}\land P^I_{\ \ s}\\
\label{2}&&R^r_{\,\ 0}=-P^{r}_{\ \ I}\land P^I_{\ \ 0}\\
\label{3}&&R^I_{\,\ J}=-P^I_{\ \ r}\land P^r_{\ \ J}-P^I_{\ \
0}\land P^0_{\ \ J}\\ \label{4}&&\nabla P^I_{\,\ r}=0\\
\label{5}&&\nabla P^I_{\,\ 0}=0
\end{eqnarray}
\noindent where
\begin{eqnarray}
&&R^{rs}\equiv d\Omega^r_{\ \ s}+\Omega^{r}_{\ \
t}\land\Omega^t_{\ \ s}+\Omega^{r}_{\ \ 0}\land\Omega^0_{\ \ s}\\
&&R^{r0}\equiv d\Omega^r_{\ \ 0}+\Omega^{r}_{\ \
t}\land\Omega^t_{\ \ 0}\\ &&R^{IJ}\equiv d\Omega^I_{\ \
J}+\Omega^{I}_{\ \ K}\land\Omega^K_{\ \ J}
\end{eqnarray}
\noindent and we have set
\begin{displaymath}
P^I_{\alpha}=\left\{ \begin{array}{rr}P^I_{\,\ 0}\equiv
\Omega^{I}_{\ \ 0}\\ P^I_{\,\ r}\equiv \Omega^{I}_{\ \
r}\end{array}\right.
\end{displaymath}
\noindent Note that $P^I_0$, $P^I_r$ are the vielbeins of the
coset, while $(\Omega^{rs},\,\ \Omega^{r0})$, $(R^{rs},\,\
R^{ro})$ are respectively the connections and the curvatures of
$SO(4)$ decomposed with respect to the diagonal subgroup
$SU(2)\subset SO(4)$.\\ In terms of the previous definitions, the
ungauged superspace curvatures of the matter coupled theory,
generalizing eqs. (\ref{T}) -(\ref{s}) (with $m=0$) are now given
by: {\setlength\arraycolsep{1pt}\begin{eqnarray}
&T^{A}&=\mathcal{D}V^{a}-\frac{i}{2}\ \
\overline{\psi}_{A}\gamma_{a}\psi^A V^{a}=0\\
&R^{ab}&=\mathcal{R}^{ab}\\
&H&=dB+2 e^{-2\sigma}\,\
\overline{\psi}_{A}\gamma_{7}\gamma_{a}\psi^AV^{a}\\
&F^{\Lambda}&=\mathcal{F}^{\Lambda}-ie^{\sigma}L^{\Lambda}_0\epsilon^{AB}
\overline{\psi}_{A}\gamma_{7}\psi_{B}-ie^{\sigma}L^{\Lambda}_r\sigma^{rAB}
\overline{\psi}_{A}\psi_{B}\\
&\rho_{A}&=\mathcal{D}\psi_{A}-\frac{i}{2}\sigma_{rAB}(-\frac{1}{2}\epsilon^{rst}\Omega_{st}-i\gamma_7\Omega_{r0})\psi^B\\
&D\chi_{A}&=\mathcal{D}\chi_{A}-\frac{i}{2}\sigma_{rAB}(-\frac{1}{2}\epsilon^{rst}\Omega_{st}-i\gamma_7\Omega_{r0})\chi^B\\
&R(\sigma)&=d\sigma\\
&\nabla\lambda_{IA}&=\mathcal{D}\lambda_{IA}-\frac{i}{2}\sigma_{rAB}(-\frac{1}{2}\epsilon^{rst}\Omega_{st}-i\gamma_7\Omega_{r0})\lambda_I^B\\
&R^{I}_0(\phi)&\equiv P^{I}_0\\
&R^{I}_r(\phi)&\equiv P^{I}_r
\end{eqnarray}}
\noindent where the last two equations define the "curvatures" of
the matter scalar fields $\phi^i$ as the vielbein of the coset:
\begin{equation}
P^{I}_0\equiv P^{I}_{0 i}d\phi^i\ \ \ \ P^{I}_r\equiv P^{I}_{r
i}d\phi^i
\end{equation}
\noindent where $i$ runs over the $4n$ values of the coset
vielbein world-components.

As in the pure supergravity case one can now write down the
superspace Bianchi identities for the matter coupled curvatures.
The computation is rather long but straightforward. We limit
ourselves to give the new transformation laws of all the physical
fields when matter is present, as derived from the solutions of
the Bianchi identities.
{\setlength\arraycolsep{1pt}\begin{eqnarray} &\delta
V^{a}_{\mu}&=-i\overline{\psi}_{A\mu}\gamma^{a}\varepsilon^A\\
&\delta B_{\mu\nu}&=2 e^{-2\sigma}
\overline{\chi}_{A}\gamma_{7}\gamma_{\mu\nu}\varepsilon^A
-4e^{-2\sigma}\overline{\varepsilon}_A\gamma_7\gamma_{[\mu}\psi_{\nu]}^A\\
 &\delta A^{\Lambda}_{\mu}&=2 e^{\sigma}
 \overline{\varepsilon}^{A}\gamma_{7}\gamma_{\mu}\chi^BL^{\Lambda}_0\epsilon_{AB}+2e^{\sigma}\overline{\varepsilon}^{A}\gamma_{\mu}\chi^{B}L^{\Lambda r}\sigma_{rAB}-e^{\sigma}L_{\Lambda
I}\overline{\varepsilon}^{A}\gamma_{\mu}\lambda^{IB}\epsilon_{AB}+\nonumber\\&&+2ie^{\sigma}L^{\Lambda}_0\overline{\varepsilon}_A\gamma^7\psi_B\epsilon^{AB}+2ie^{\sigma}L^{\Lambda r}\sigma_{r}^{AB}\overline{\varepsilon}_A\psi_B\\
\label{qui}&\delta\psi_{A\mu}&=\mathcal{D}_{\mu}\varepsilon_A+\frac{1}{16}
e^{-\sigma}[T_{[AB]\nu\lambda}\gamma_{7}-T_{(AB)\nu\lambda}](\gamma_{\mu}^{\,\
\nu\lambda}-6\delta_{\mu}^{\nu}\gamma^{\lambda})
\varepsilon^{B}+\nonumber \\
 &&+\frac{i}{32}e^{2\sigma} H_{\nu\lambda\rho}
\gamma_{7}(\gamma_{\mu}^{\,\ \nu\lambda\rho}-3\delta_{\mu}^{\nu}
\gamma^{\lambda\rho})\varepsilon_{A}+\frac{1}{2}\varepsilon_{A}\overline{\chi}^{C}\psi_{C\mu}+\nonumber\\
&&+\frac{1}{2}\gamma_{7}\varepsilon_{A}\overline{\chi}^{C}\gamma^{7}\psi_{C\mu}-\gamma_{\nu}\varepsilon_{A}\overline{\chi}^{C}\gamma^{\nu}\psi_{C\mu}+\gamma_{7}\gamma_{\nu}\varepsilon_{A}\overline{\chi}^{C}\gamma^{7}\gamma^{\nu}\psi_{C\mu}+\nonumber\\
&&-\frac{1}{4}\gamma_{\nu\lambda}\varepsilon_{A}\overline{\chi}^{C}\gamma^{\nu\lambda}\psi_{C\mu}-\frac{1}{4}\gamma_{7}\gamma_{\nu\lambda}\varepsilon_{A}\overline{\chi}^{C}\gamma^{7}\gamma^{\nu\lambda}\psi_{C\mu}\\
\label{quo}&\delta\chi_{A}&=\frac{i}{2}
\gamma^{\mu}\partial_{\mu}\sigma \varepsilon_{A}\!+\!
\frac{i}{16}e^{-\sigma}[T_{[AB]\mu\nu}\gamma_{7}\!+\!T_{(AB)\mu\nu}]\gamma^{\mu\nu}\varepsilon^{B}\!+\!\frac{1}{32}e^{2\sigma}
H_{\mu\nu\lambda}\gamma_{7}\gamma^{\mu\nu\lambda}\varepsilon_{A}\\
&\delta\sigma&=\overline{\chi}_{A}\varepsilon^A\\
\label{qua}&\delta\lambda^{IA}&=-iP^I_{ri}\sigma^{rAB}\partial_{\mu}\phi^{i}\gamma^{\mu}\varepsilon_{B}+iP^I_{0i}\epsilon^{AB}\partial_{\mu}\phi^{i}\gamma^{7}\gamma^{\mu}\varepsilon_{B}+\frac{i}{2}e^{-\sigma}T^{I}_{\mu\nu}\gamma^{\mu\nu}\varepsilon^{A}\\
&P^{I}_{0i}\delta\phi^i&=\frac{1}{2}\overline{\lambda}^{I}_{A}\gamma_{7}\varepsilon^A\\
&P^{I}_{ri}\delta\phi^i&=\frac{1}{2}\overline{\lambda}^{I}_{A}\varepsilon_{B}\sigma_r^{ab}
\end{eqnarray}}
\noindent where we have introduced the "dressed" vector field
strengths
\begin{eqnarray}
&&T_{[AB]\mu\nu}\equiv\epsilon_{AB}L^{-1}_{0\Lambda
}F^{\Lambda}_{\mu\nu}\\
&&T_{(AB)\mu\nu}\equiv\sigma^r_{AB}L^{-1}_{r\Lambda}F^{\Lambda}_{\mu\nu}\\
&&T_{I\mu\nu}\equiv L^{-1}_{I\Lambda }F^{\Lambda}_{\mu\nu}
\end{eqnarray}
\noindent and we have omitted in the transformation laws of the fermions the three-fermions terms of the form $(\chi\chi\varepsilon)$, $(\lambda\lambda\varepsilon)$.\\
We observe that the solutions of the Bianchi identities
also imply the equations of motion of the physical fields and
therefore one can reconstruct in principle the space-time
Lagrangian.  Nevertheless, in general, it is simpler to construct
the Lagrangian explicitly and we will present its complete
expression in the forthcoming paper of reference \cite{forth}. In
the next section, however, we will need at least the kinetic
terms and mass matrices terms in order to construct the scalar
potential of the gauged theory. This is the topic of the next
paragraph.

\section{The gauging}
The next problem we have to cope with is the gauging of the
matter coupled theory and the determination of the scalar
potential.\\
Let us first consider the ordinary gauging, with $m=0$, which, as
usual, will imply the presence of new terms proportional to the
coupling constants in the supersymmetry transformation laws of
the fermion fields.\\
Our aim is to gauge a compact subgroup of $O(4,n)$. Since in any
case we may gauge only the diagonal subgroup $SU(2)\subset
O(4)\subset H$, the maximal gauging is given by
$SU(2)\otimes\mathcal{G}$ where $\mathcal{G}$ is a $n$-dimensional
subgroup of $O(n)$. According to a well known procedure, we
modify the definition of the left invariant 1-form $L^{-1}dL$ by
replacing the ordinary differential with the
$SU(2)\otimes\mathcal{G}$ covariant differential as follows:
\begin{equation}
\label{nabla}\nabla L^{\Lambda}_{\ \ \Sigma}=d L^{\Lambda}_{\ \
\Sigma}-f_{\Gamma\ \ \Pi}^{\,\ \Lambda} A^{\Gamma} L^{\Pi}_{\ \
\Sigma}
\end{equation}
\noindent where $f^{\Lambda}_{\ \ \Pi\Gamma}$ are the structure
constants of $SU(2)_d\otimes\mathcal{G}$. More explicitly,
denoting with $\epsilon^{rst}$ and $\mathcal{C}^{IJK}$ the
structure constants of the two factors $SU(2)$ and $\mathcal{G}$,
equation (\ref{nabla}) splits as follows:
\begin{eqnarray}
&&\nabla L^{0}_{\ \ \Sigma}=d L^{\Lambda}_{\ \ \Sigma}\\
&&\nabla L^{r}_{\ \ \Sigma}=d L^{r}_{\ \ \Sigma}-g\epsilon^{\,\
r}_{t\
\ s} A^{t} L^{s}_{\ \ \Sigma}\\
&&\nabla L^{I}_{\ \ \Sigma}=d L^{I}_{\ \
\Sigma}-g'\mathcal{C}^{\,\ I}_{K\ \ J} A^{K} L^{J}_{\ \ \Sigma}
\end{eqnarray}
\noindent Setting $\widehat{\Omega}=L^{-1}\nabla L$, one easily
obtains the gauged Maurer-Cartan equations:
\begin{equation}\label{mc}d\widehat{\Omega}^{\Lambda}_{\ \
\Sigma}+\widehat{\Omega}^{\Lambda}_{\ \
\Pi}\land\widehat{\Omega}^{\Pi}_{\ \
\Sigma}=(L^{-1}\mathcal{F}L)^{\Lambda}_{\ \ \Sigma}
\end{equation}
\noindent where $\cF\equiv\cF^{\Lambda}T_{\Lambda}$, $T_{\Lambda}$ being the generators of $SU(2)\otimes\cG$.\\
After gauging, the same decomposition as in eqs. (\ref{1}) -(\ref{5}) now gives:
\begin{eqnarray}
\label{11}&&R^r_{\,\ s}=-P^{r}_{\ \ I}\land P^I_{\ \ s}+(L^{-1}\mathcal{F}L)^{r}_{\ \ s}\\
\label{12}&&R^r_{\,\ 0}=-P^{r}_{\ \ I}\land P^I_{\ \ 0}+(L^{-1}\mathcal{F}L)^{r}_{\ \ 0}\\
\label{13}&&R^I_{\,\ J}=-P^I_{\ \ r}\land P^r_{\ \ J}-P^I_{\ \
0}\land P^0_{\ \ J}+(L^{-1}\mathcal{F}L)^{I}_{\ \ J}\\ \label{14}&&\nabla P^I_{\,\ r}=(L^{-1}\mathcal{F}L)^{I}_{\ \ r}\\
\label{15}&&\nabla P^I_{\,\ 0}=(L^{-1}\mathcal{F}L)^{I}_{\ \ 0}
\end{eqnarray}
Because of the presence of the gauged terms in the coset
curvatures, the new Bianchi Identities are not satisfied by the
old superspace curvatures but we need extra terms in the fermion
field strengths parametrizations, that is, in space-time language,
extra terms in the transformation laws of the fermion fields of
eqs. (\ref{qui}), (\ref{quo}), (\ref{qua}), named ``fermionic shifts''.
\begin{eqnarray}
\label{del1}&&\delta\psi_{A\mu}=\delta\psi_{A\mu}^{(old)}+S_{AB}(g,g')\gamma_{\mu}\varepsilon^B\\
\label{del2}&&\delta\chi_A=\delta\chi_A^{(old)}+N_{AB}(g,g')\varepsilon^B\\
\label{del3}&&\delta\lambda_A^I=\delta\lambda^{I
(old)}_A+M^I_{AB}(g,g')\varepsilon^B
\end{eqnarray}
\noindent Again, working out the Bianchi identities, one fixes the
explicit form of the fermionic shifts which turn out to be
\begin{eqnarray}
\label{S}&&S_{AB}^{(g,g')}=\frac{i}{24}Ae^{\sigma}\epsilon_{AB}-\frac{i}{8}B_t\gamma^7\sigma^t_{AB}\\
\label{N}&&N_{AB}^{(g,g')}=\frac{1}{24}Ae^{\sigma}\epsilon_{AB}+\frac{1}{8}B_t\gamma^7\sigma^t_{AB}\\
\label{M}&&M^{I(g,g')}_{AB}=(-C^I_t+2i\gamma^7D^I_t)\sigma^t_{AB}
\end{eqnarray}
\noindent where
\begin{eqnarray}\label{AA}&&A=\epsilon^{rst}K_{rst}\\  \label{BB}&&B^i=\epsilon^{ijk}K_{jk0}\\ \label{CC}&&C_I^t=\epsilon^{trs}K_{rIs}\\ \label{DD}&&D_{It}=K_{0It}\end{eqnarray}
\noindent and the threefold completely antisymmetric tensors
$K's$ are the so called "boosted structure constants" given
explicitly by:
\begin{eqnarray}
&&K_{rst}=g\epsilon_{lmn}L^l_{\,\ r}(L^{-1})^{\,\ m}_sL^n_{\,\
t}+g'\mathcal{C}_{IJK}L^I_{\,\ r}(L^{-1})^{\,\ J}_sL^K_{\,\ t}\\
&&K_{rs0}=g\epsilon_{lmn}L^l_{\,\ r}(L^{-1})^{\,\ m}_sL^n_{\,\
0}+g'\mathcal{C}_{IJK}L^I_{\,\ r}(L^{-1})^{\,\ J}_sL^K_{\,\ 0}\\
&&K_{rIt}=g\epsilon_{lmn}L^l_{\,\ r}(L^{-1})^{\,\ m}_IL^n_{\,\
t}+g'\mathcal{C}_{LJK}L^L_{\,\ r}(L^{-1})^{\,\ J}_IL^K_{\,\ t}\\
&&K_{0It}=g\epsilon_{lmn}L^l_{\,\ 0}(L^{-1})^{\,\ m}_IL^n_{\,\
t}+g'\mathcal{C}_{LJK}L^L_{\,\ 0}(L^{-1})^{\,\ J}_IL^K_{\,\ t}
\end{eqnarray}
Actually one easily see that the fermionic shifts (\ref{S})
(\ref{N}) reduce to the pure supergravity $g$ dependent terms of
equations (\ref{obiwan}), (\ref{mfw}). (Note that, since
$L^{\Lambda}_{\ \ \Sigma}\rightarrow\delta^{\Lambda}_{\Sigma}$ in
absence of matter, the terms proportional to the Pauli $\sigma$
matrices are simply absent in such a limit.)\\
At this point one could compute the scalar potential of the
matter coupled theory, in terms of the fermionic shifts just
determined, using the well known Ward identity of the scalar
potential which can be derived from the Lagrangian. Since we are
going to perform this derivation once we will introduce also $m$
dependent terms in the fermionic shifts, we just quote, for the
moment the expected result that the potential due only to $g$ and
$g'$ dependent shifts doesn't admit a stable $AdS$ background
configuration. We are thus, led as in the pure supergravity case,
to determine suitable $m$ dependent terms that reduce to the $m$
terms of eqs. (\ref{bart}), (\ref{homer}) in absence of matter
multiplets. One can see that a simple-minded ansatz of keeping
exactly the same form for the $m$ dependent terms as in the pure
Supergravity case is not consistent with the gauged superspace
Bianchi identities.\\
It turns out that a consistent solution for the relevant $m$ terms
to be added to the fermionic shifts, implies the presence of the
coset representatives; that is, the $m$-terms must also be
"dressed" with matter scalar fields as it happens for the $g$ and
$g'$ dependent terms. Explicitly, the Bianchi identities solution
for the new fermionic shifts is:
\begin{eqnarray}
\label{mS}&&
S_{AB}^{(g,g',m)}=\frac{i}{24}[Ae^{\sigma}+6me^{-3\sigma}(L^{-1})_{00})\epsilon_{AB}-\frac{i}{8}[B_te^{\sigma}-2me^{-3\sigma}(L^{-1})_{i0}]\gamma^7\sigma^t_{AB}\\
\label{mN}&&N_{AB}^{(g,g',m)}\!\!=\!\!\frac{1}{24}[Ae^{\sigma}-18me^{-3\sigma}(L^{-1})_{00})]\epsilon_{AB}+\frac{1}{8}[B_te^{\sigma}+6me^{-3\sigma}(L^{-1})_{i0}]\gamma^7\sigma^t_{AB}\\
\label{mM}&&M^{I(g,g',m)}_{AB}=(-C^I_t+2i\gamma^7D^I_t)e^{\sigma}\sigma^t_{AB}-2me^{-3\sigma}(L^{-1})^I_{\
\ 0}\epsilon_{AB}
\end{eqnarray}
Let us note that, similarly to what happens in the pure
supergravity case, the $m$ dependent terms behave in an analogous way as
the $g$ dependent gauging  terms, the former being dressed by the
scalar fields in a similar way as the latters. The difference is that, while the gauged terms are
threelinear in the coset representatives of the duality group,
the $m$ terms are linear (recall that because of the
pseudo-orthogonality condition, $L^{-1}=\eta
L^T\eta$).\footnote{Amusingly enough the reverse situation happens
for the "coset" representative of $O(1,1)$: the $g$ terms are
linear in $e^{\sigma}$, while the $m$ terms are threelinear, being
proportional to $e^{-3\sigma}$.} Furthermore we note that, as it
happens for the $g$ and $g'$ terms, the matter coupling forces
new terms in the fermionic shifts proportional to $\sigma^t_{AB}$
while for the gaugino shift the $m$ terms contribute terms
proportional to $\epsilon_{AB}$.

\section{The scalar potential}
The simplest way to derive the scalar potential is to use the
supersymmetry Ward identity which relates the scalar potential to
the fermionic shifts in the transformation laws \cite{fema}. In order to
retrieve such identity it is necessary to have the relevant terms
of the Lagrangian of the gauged theory. These terms are actually
the kinetic ones and the "mass" terms given in the following
equation:
\begin{eqnarray}
\label{lag}&&(detV)^{-1}\mathcal{L}=-\frac{1}{4}\mathcal{R}
-\frac{1}{8}e^{2\sigma}\mathcal{N}_{\Lambda\Sigma}\widehat{\mathcal{F}}^{\Lambda}_{\mu\nu}\widehat{\mathcal{F}}^{\Sigma\mu\nu}+\partial^{\mu}\sigma\partial_{\mu}\sigma
-\frac{1}{4}(P^{I0}_{\mu}P^{\mu}_{I0}
+P^{Ir}_{\mu}P^{\mu}_{Ir})+\nonumber\\
&&-\frac{i}{2}\overline{\psi}_{A\mu}\gamma^{\mu\nu\rho}D_{\nu}\psi^{A}_{\rho}+\frac{i}{8}\overline{\lambda}^I_A\gamma^{\mu}D_{\mu}\lambda^A_I-2i\overline{\chi}_A\gamma^{\mu}D_{\mu}\chi^A
+2i\overline{\psi}_{\mu}^A\gamma^{\mu\nu}\overline{S}_{AB}\psi_{\nu}^B+\nonumber\\
&&+4i\overline{\psi}_{\mu}^A\gamma^{\mu}\overline{N}_{AB}\chi^{B}+\frac{i}{4}\overline{\psi}_{\mu}^A\gamma^{\mu}\overline{M}_{AB}^I\lambda^{B}_I+\mathcal{W}(\sigma\phi^i;g,g',m)+\dots
\end{eqnarray}
\noindent where
\begin{equation}
\label{kin}\mathcal{N}_{\Lambda\Sigma}=L_{\Lambda}^{\,\
0}L^{-1}_{0\Sigma}+L_{\Lambda}^{\,\
i}L^{-1}_{i\Sigma}-L_{\Lambda}^{\,\ I}L^{-1}_{I\Sigma}
\end{equation}
\noindent is the vector kinetic matrix,
$\widehat{\mathcal{F}}^{\Lambda}_{\mu\nu}\equiv\mathcal{F}^{\Lambda}_{\mu\nu}-m\delta^{\Lambda}_0B_{\mu\nu}$ and $\cW$ is minus the scalar potential.\\
 In equation
(\ref{lag}) there appear ``barred mass-matrices'' $\overline{S}_{AB},\,\
\overline{N}_{AB},\,\ \overline{M}^I_{AB}$ which are
slightly different from the fermionic shifts defined in eqs.
(\ref{mS}), (\ref{mN}), (\ref{mM}). Actually they are defined by:
\begin{equation}
\label{pippo}\overline{S}_{AB}=-S_{BA},\ \ \ \
\overline{N}_{AB}=-N_{BA},\ \ \ \ \overline{M}^I_{AB}=M^I_{BA}
\end{equation}
\noindent Definitions (\ref{pippo}) stem from the fact
that the shifts defined in eqs. (\ref{mS}), (\ref{mN}),
(\ref{mM}) are matrices in the eight-dimensional spinor space,
since they contain the $\gamma_7$ matrix; as will be seen in a
moment, such definition is actually necessary in order to satisfy the
supersymmetry Ward identity.\\
 Indeed, let us perform the
supersymmetry variation of (\ref{lag}), keeping only the terms
proportional to $g$, $g'$ or $m$, and to the current
$\overline{\psi}_{A\mu}\gamma^{\mu}\epsilon^A$; we find the
following Ward identity, :
\begin{equation}
\label{ward}\delta^C_A\mathcal{W}=20\overline{S}^{AB}S_{BC}+4\overline{N}^{AB}N_{BC}+\frac{1}{4}\overline{M}^{AB}_IM^I_{BC}
\end{equation}
\noindent However we note that, performing the supersymmetry
variation, the gauge terms also give  rise to extra terms
proportional to the current
$\overline{\psi}_{A\mu}\gamma^7\gamma^{\mu}\epsilon^A$, which
have no counterpart in the term containing the potential $\cW$. Because
of the definition of the barred mass matrices in eq.
(\ref{pippo}) it is easily seen that such "$\gamma^7$-terms",
arising from $\overline{S}^{AB}S_{BC}$ and
$\overline{N}^{AB}N_{BC}$ cancel
against each other.\\
As far as the term $\overline{M}^{AB}_IM^I_{BC}$ is concerned,
 the same mechanism of cancellation again applies to the terms
proportional to
$\overline{\psi}_{A\mu}\gamma_7\gamma^{\mu}\epsilon^A\sigma^{rA}_C$;
there is, however, a residual dangerous term of the form
\begin{equation}
\delta^A_C\overline{\psi}_{A\mu}\gamma^{\mu}\gamma^7D^I_{\,\
s}C_I^{\,\ s}\epsilon^C
\end{equation}
\noindent One can show that this term vanishes identically owing
to the non trivial relation
\begin{equation}
\label{abigaille}D^I_tC_I^t=0
\end{equation}
\noindent Equation (\ref{abigaille}) can be shown to hold  using
the pseudo-orthogonality relation $L^T\eta L=\eta$ among the coset
representatives and the Jacobi identities $C_{I[JK}C_{L]MN}=0$,
$\epsilon_{r[st}\epsilon_{l]mn}=0$. This is a non trivial check of our computation.\\
It now follows that the Ward identity eq. (\ref{ward}) is indeed
satisfied since all the terms on the r.h.s., once the "$\gamma^7$-terms" have been cancelled,  are proportional to $\delta^C_A$.\\
Using the expressions (\ref{mS}), (\ref{mN}), (\ref{mM}),
(\ref{pippo}) in equation (\ref{ward}), we obtain the explicit
form of the scalar potential
{\setlength\arraycolsep{1pt}\begin{eqnarray}\label{pot}\mathcal{W}(\phi)=&&
5\,\ \{
[\frac{1}{12}(Ae^{\sigma}+6me^{-3\sigma}L_{00})]^2+[\frac{1}{4}(e^{\sigma}B_i-2me^{-3\sigma}L_{0i})]^2\}+\nonumber\\
\nonumber\\ &&- \{
[\frac{1}{12}(Ae^{\sigma}-18me^{-3\sigma}L_{00})]^2+[\frac{1}{4}(e^{\sigma}B_i+6me^{-3\sigma}L_{0i})]^2\}+\nonumber\\
&&-\frac{1}{4}\{C^I_{\,\ t}C_{It}+4D^I_{\,\ t}D_{It}\}\,\
e^{2\sigma}-m^2e^{-6\sigma}L_{0I}L^{0I}
\end{eqnarray}}
\noindent Let us note that, setting
\begin{eqnarray}
&&\mathcal{H}=\frac{1}{12}(Ae^{\sigma}+6me^{-3\sigma}L_{00})\\
&&\mathcal{K}_i=\frac{1}{4}(e^{\sigma}B_i+6me^{-3\sigma}L_{0i})
\end{eqnarray}
\noindent the potential can be written as follows
\begin{eqnarray}
&&\mathcal{W}=5\{\mathcal{H}^2+\mathcal{K}^i\mathcal{K}_i\}-\{[\partial_{\sigma}\mathcal{H}]^2+\partial_{\sigma}\mathcal{K}^i\nabla_{\sigma}\mathcal{K}_i\}-2\{\nabla_{I\alpha}\mathcal{H}\nabla^{I\alpha}\mathcal{H}+\nabla_{I\alpha}\mathcal{K}_i\nabla^{I\alpha}\mathcal{K}^i\}_{m=0}+\nonumber\\
\label{bibi}&&-\{\nabla_{I\alpha}\mathcal{H}\nabla^{I\alpha}\mathcal{H}+\nabla_{I\alpha}\mathcal{K}_i\nabla^{I\alpha}\mathcal{K}^i\}_{g=0}
\end{eqnarray}
\noindent or alternatively
\begin{eqnarray}
\label{bibo}&&\mathcal{W}=5\{\mathcal{H}^2+\mathcal{K}^i\mathcal{K}_i\}-\{[\partial_{\sigma}\mathcal{H}]^2+\partial_{\sigma}\mathcal{K}^i\partial_{\sigma}\mathcal{K}_i\}
-2\{\nabla_{I\alpha}\mathcal{H}\nabla^{I\alpha}\mathcal{H}+\nabla_{I\alpha}\mathcal{K}_i\nabla^{I\alpha}\mathcal{K}^i\}+\nonumber\\
&&+m^2e^{-6\sigma}L_{0I}L^{0I}
\end{eqnarray}
\noindent where $\nabla_{I\alpha}\equiv
(\nabla_{I0},\nabla_{Ir})$ denote the derivatives with respect
to the "linearized coordinates": that is, using the Maurer-Cartan
equations
\begin{eqnarray}
&&\nabla^H L^{\Lambda}_{\ \ I}=L^{\Lambda}_{\ \
\alpha}P^{\alpha}_I\nonumber\\ \label{ugo}&&\nabla^H
L^{\Lambda}_{\ \ \alpha}=L^{\Lambda}_{\ \ I}P_{\alpha}^I
\end{eqnarray}
\noindent the flat derivative $\nabla_{I\alpha}$ are defined as
the coefficient of the coset vielbein $P^{I\alpha}$ in equations
(\ref{ugo}). In deriving equations (\ref{bibi}), (\ref{bibo}) one
has to make use of the following relations which are a
straightforward consequence of the definitions (\ref{AA})
-(\ref{DD})
\begin{eqnarray}
&&\nabla_{I0}A=0\\
&&\nabla_{Ir}A=C_{Ir}\\
&&\nabla_{I0}B_i=C_{iI}\\
&&\nabla_{Ir}B_i=2\epsilon_{rik}D_{Ik}
\end{eqnarray}
\noindent Expanding the squares in equation (\ref{pot})the
potential $\mathcal W$can be alternatively written as follows:
\begin{eqnarray}
&&\mathcal{W}=e^{2\sigma}[\frac{1}{36}A^2+\frac{1}{4}B^iB_i-\frac{1}{4}(C^I_{\,\
t}C_{It}+4D^I_{\,\
t}D_{It})]-m^2e^{-6\sigma}\mathcal{N}_{00}+\nonumber\\
&&+me^{-2\sigma}[\frac{2}{3}AL_{00}-2B^iL_{0i}]
\end{eqnarray}
\noindent where $\mathcal{N}_{00}$ is the 00 component of the
vector kinetic matrix defined in eq. (\ref{kin}).\\
We now show
that, apart from other possible extrema not considered here, a stable supersymmetric extremum of the potential
$\mathcal{W}$ is found to be the same as in the case of pure supergravity,
that is we get an $AdS$ supersymmetric background only for
$g=3m$. In fact, setting $\partial_{\sigma}\mathcal{W}=0$ and keeping
only the non vanishing terms at $\sigma=q^{I\alpha}=0$, $q^{I\alpha}$
being the flat coordinates, we have
\begin{eqnarray}
&&\partial_{\sigma}\mathcal{W}=[\frac{1}{18}A^2e^{2\sigma}-\frac{4}{3}mAL_{00}e^{-2\sigma}+6m^2L_{00}^2e^{-6\sigma}]_{\sigma=q^{I\alpha}=0}
\end{eqnarray}
\noindent since all the other terms entering the $\partial_{\sigma}\cW$ contain at least one
off-diagonal element of the coset representative which vanishes
identically when the scalar fields are set equal to zero.
Furthermore, from the definition (\ref{AA}) and using
$L^{\Lambda}_{\,\
\Sigma}(q^I_{\alpha}=0)=\delta^{\Lambda}_{\Sigma}$, we find:
\begin{equation}\label{aq}A(q^I_{\alpha}=0)=6g;\ \ \ \ L_{00}(q^I_{\alpha}=0)=1\end{equation}
\noindent so that
\begin{eqnarray}
\label{pluto}&&\partial_{\sigma}\mathcal{W}|_{\sigma=q=0}=2g^2-8mg+6m^2=0
\end{eqnarray}
\noindent  As the partial derivatives
$(\frac{\partial\mathcal{W}}{\partial q^{I0}})_{\sigma=q=0}$,
$(\frac{\partial\mathcal{W}}{\partial q^{Ir}})_{\sigma=q=0}$ are
also zero, since they contains at least one off-diagonal coset
representative, the condition for the minimum is given by eq.
(\ref{pluto}) which coincides with the equation one obtains for
the pure Supergravity case, whose solutions are $g=m$, $g=3m$.\\
Using equations (\ref{aq}), (\ref{pluto}), (\ref{del1}) -(\ref{del3}) one  can  easily recognize  that only
the $g=3m$ solution gives rise to a supersymmetric $AdS$ background.\\
A further issue related to the scalar potential, which is an
important check of all our calculation, is the possibility of
computing the masses of the scalar fields by varying the
linearized kinetic terms and the potential of (\ref{lag}), after
power expansion of $\mathcal{W}$ up to the second order in the
scalar fields $q^I_{\alpha}$.\\
We find:
\begin{eqnarray}
&&(\frac{\partial^2\mathcal{W}}{\partial\sigma^2})_{\sigma=q=0,
g=3m}=48m^2\\
&&(\frac{\partial^2\mathcal{W}}{\partial q^{I0}\partial q^{J0}}
)_{\sigma=q=0,
g=3m}=8m^2\delta^{IJ}\\
&&(\frac{\partial^2\mathcal{W}}{\partial q^{Ir}\partial q^{Js}
})_{\sigma=q=0, g=3m}=24m^2\delta^{IJ}\delta^{rs}
\end{eqnarray}
\noindent The linearized equations of motion become:
\begin{eqnarray}
&&\Box\sigma-24m^2\sigma=0\\
&&\Box q^{I0}-16m^2 q^{I0}=0\\
&&\Box q^{Ir}-24m^2 q^{Ir}=0
\end{eqnarray}
\noindent If we use as mass unity the
\def\IP{\relax{\rm I\kern-.18em P}} inverse $AdS$ radius, which
in our conventions (see eq.(\ref{curv})) is $R^{-2}_{AdS}=4m^2$ we get:
\begin{eqnarray}
&&m^2_{\sigma}=-6\nonumber\\
&&m^2_{q^{I0}}=-4\nonumber\\
\label{massa}&&m^2_{q^{Ir}}=-6
\end{eqnarray}
\noindent These values should be compared with the results
obtained in reference \cite{fkpz} where the supergravity and
matter multiplets of the $AdS_6\,\ F(4)$ theory were constructed
in terms of the singleton fields of the 5-dimensional conformal
field theory, the singleton being given by hypermultiplets
transforming in the fundamental of $\mathcal{G}\equiv E_7$. It is
amusing to see that the values of the masses of the scalars
computed in terms of the conformal dimensions are exactly the
same as those given in equation (\ref{massa}).\\ This coincidence
can be considered as a non trivial check of the $AdS/CFT$
correspondence in six versus five dimensions.\\
To make contact with what follows we observe that the scalar squares masses in $AdS_{d+1}$ are given by the $SO(2,d)$ quadratic Casimir \cite{flafro}
\be
m^2=E_0(E_0-d)
\ee
\noindent They are negative in the interval $\frac{d-2}{2}\leq E_0<d$ (the lower bound corresponding to the unitarity bound i.e. the singleton) and attain the Breitenlohner-Freedman bound \cite{breit} when $E_0=d-E_0$ i.e. at $E_0=\frac{d}{2}$ for which $m^2=-\frac{d^2}{4}$. Conformal propagation correspond to $m^2=-\frac{d^2-1}{4}$ i.e. $E_0=\frac{d\pm 1}{2}$. This is the case of the dilaton and triplet matter scalars.

\section{$F(4)\otimes\cG$ Superconformal Field Theory}
In this section we describe the basics of the  $F(4)$ highest
weight unitary irreducible representations ``UIR's'' and exhibit
two towers of short representations which are relevant for a K-K
analysis of type IIA theory on  (warped) $AdS_6\otimes S^4$
\cite{oz},\cite{clp}.\\ We will not consider here the $\cG$
representation properties but we will only concentrate on the
supersymmetric structure.\\ Recalling that the even part of the
$F(4)$ superalgebra is $SO(2,5)\otimes SU(2)$, from a general
result on Harish-Chandra modules \cite{f},\cite{min}, \cite{g},
\cite{h} of $SO(2,2n+1)$ we know that there are only a spin 0 and
a spin 1/2 singleton unitary irreducible representations
\cite{flafro}, which, therefore, merge into a unique
supersingleton representation of the $F(4)$ superalgebra: the
hypermultiplet \cite{fkpz}.\\ To describe shortening is useful to
use a harmonic superfield language \cite{fanta1}.\\ The harmonic
space is in this case the 2-sphere\footnote{The sphere is the
simplest example  of ``flag manifold'' whose geometric structure
underlies the construction of harmonic superspaces \cite{hh}}
$SU(2)/U(1)$, as in $N=2,\,\ d=4$ and $N=1,\,\ d=6$. A highest
weight UIR of $SO(2,5)$ is determined by $E_0$ and a UIR of
$SO(5)\simeq Usp(4)$, with Dynkin labels $(a_1,\,\ a_2)$
\footnote {Note that the $Usp(4)$ Young labels $h_1,h_2$ are
related to $a_1,a_2$ by $a_1=2h_2; a_2=h_1-h_2$.}. We will denote
such representations by $\cD(E_0,\,\ a_1,\,\ a_2)$. The two
singletons correspond to $E_0=3/2$, $a_1=a_2=0$ and $E_0=2$,
$a_1=1$, $a_2=0$.\\ In the $AdS/CFT$ correspondence $(E_0,\,\
a_1,\,\ a_2)$ become the conformal
dimension and the Dynkin labels of $SO(1,4)\simeq Usp(2,2)$.\\
The highest weight UIR of the $F(4)$ superalgebra will be denoted
by $\cD(E_0,\,\ a_1,\,\ a_2;\,\ I)$ where $I$ is the $SU(2)$
$R$-symmetry quantum number (integer or half integer).\\ We will
show shortly that there are two (isolated) series of UIR's which
correspond respectively to $1/2$ BPS short multiplets (analytic
superfields) and intermediate short superfields. The former have
the property that they form a ring under multiplication, as the
chiral fields in $d=4$ \cite{fesoca}.\\ The first series is the
massive
 tower of short vector multiplets  whose lowest
members is a massless vector multiplet in $Adj\cG$ corresponding
to the conserved currents of the  $\cG$ global symmetry of the
five dimensional
conformal field theory.\\
The other series is the tower of massive graviton multiplets, which
exhibit "intermediate shortening" and it is not of BPS type. Its
lowest member is the supergravity multiplet which contains the $SU(2)$
$R$-symmetry current and the stress-tensor among the  superfield components.

\subsection{$F(4)$ superfields}

The basic superfield is the supersingleton hypermultiplet
$W^A(x,\theta)$, which satisfies the constraint
\begin{equation}\label{hyper}
D_{\alpha}^{(A}W^{B)}(x,\theta)=0
\end{equation}
\noindent corresponding to the irrep. $\cD(E_0=\frac{3}{2},0,0;I=\frac{1}{2})$ \cite{fanta1}.\\
\noindent By using harmonic superspace, $(x,\,\ \theta_I,\,\
u^I_i)$, where $\theta_I=\theta_iu^i_I$, $u^i_I$ is the coset
representative of $SU(2)/U(1)$ and $I$ is the charge $U(1)$-label,
from  the covariant derivative algebra \be
\{D^A_{\alpha},D^jB_{\beta}\}=i\epsilon^{AB}\partial_{\alpha\beta}
\ee
 \noindent we have
\be \{D^I_{\alpha},D^I_{\beta}\}=0 \ \ \ \
D^I_{\alpha}=D^i_{\alpha}u^I_i \ee \noindent Therefore from eq.
(\ref{hyper}) it follows the $G$-analytic constraint: \be
D_{\alpha}^{1}W^{1}=0\ee \noindent which implies \be
W^1(x,\theta)=\varphi^1+\theta^{\alpha}_2\zeta_{\alpha}+d.t. \ee
 \noindent (d.t. means ``derivative terms'').\\
Note that $W^1$ also satisfies \be D^2_{\alpha}D^{2\alpha}W^1=0
\ee
 \noindent because there is no such scalar component\footnote{This is rather similar to the treatment of the (1,0) hypermultiplet in $D=6$ \cite{fanta2}} in $W^1$.\\
$W^1$ is a Grassman  analytic  superfield, which is also harmonic
(that is ${\bf D}^1_2W^1=0$ where, using notations of reference \cite{fesoca}, ${\bf D}^1_2$ is the step-up operator
of the
$SU(2)$ algebra acting on  harmonic superspace).\\ Since $W^1$ satifies $D^1W^1=0$, any $p$-order
polynomial
\be
\label{wp}I_p(W^1)=(W^1)^p\ee \noindent will also have the same property, so
these operators form a ring under multiplication \cite{fesoca}, they are the 1/2
BPS states of the $F(4)$ superalgebra and represent massive vector
multiplets $(p>2)$, and massless bulk gauge fields for $p=2$.\\
The above multiplets correspond to the $D(E_0=3I,0,0;I=\frac{p}{2})$ h.w. U.I.R.'s of the $F(4)$ superalgebra.\\
Note also that if $W^1$ carries a pseudo-real representation of
the flavor group $\cG$ (e.g. {\bf 56} of $\cG=E_7$) then $W^1$
satisfies a reality condition
\be
(W^1)^*=W^2 \ee
 \noindent corresponding to the superfield constraint
\be
(W^A)^{*\Lambda}=\epsilon_{AB}\Omega_{\Lambda\Sigma}W^{jB\Sigma}
\ee
 \noindent The $SU(2)$ quantum numbers of the $W^{1p}$ superfield
 components are:
\begin{eqnarray*} &&(\theta)^0\hspace{19 mm}spin\,\ 0\hspace{20
mm}I=\frac{p}{2}\\ &&(\theta)^1\hspace{19
mm}spin\frac{1}{2}\hspace{20 mm}I=\frac{p}{2}-\frac{1}{2}\\
&&(\theta)^2\hspace{10 mm}spin\,\ 0 - spin\,\ 1\hspace{10
mm}I=\frac{p}{2}-1\\ &&(\theta)^3\hspace{19
mm}spin\frac{1}{2}\hspace{20 mm}I=\frac{p}{2}-\frac{3}{2}\\
&&(\theta)^4\hspace{19 mm}spin\,\ 0\hspace{20 mm}I=\frac{p}{2}-2
\end{eqnarray*}
\noindent Note that the $(\theta)^4$ component is missing for  $p=2$, $p=3$, while the $(\theta)^3$ component is missing for $p=2$.
However the total number of states is $8(p-1)$ both for boson and fermion fields ($p\geq 2$).\\
 The AdS squared mass for scalars is
\be
m_s^2=E_0(E_0-5) \ee
 \noindent so there are three families of scalar states with
\begin{eqnarray*}
&&m^2_1=\frac{3}{4}p(3p-10)\hspace{20 mm}p\geq 2\\
&&m^2_2=\frac{1}{4}(3p+2)(3p-8)\hspace{10 mm}p\geq 2\\
&&m^2_3=\frac{1}{4}(3p+4)(3p-6)\hspace{10 mm}p\geq 4\\
\end{eqnarray*}
\noindent The only scalars states with $m^2<0$ are the scalar in
the massless vector multiplet $(p=2)$ with $m^2_1=-6$, $m^2_2=-4$
(no states with $m^2=0$ exist) and in the $p=3$ multiplet with $m^2=-\frac{9}{4}$.\\ We now consider the second
"short" tower containing the graviton supermultiplet and its
recurrences.\\ The graviton multiplet is given by $W^1\overline{W}^1$. Note that such superfield is not
$G$-analytic, but it satisfies
\be
D^1_{\alpha}D^{1\alpha}(W^1\overline{W}^1)=D^2_{\alpha}D^{2\alpha}(W^1\overline{W}^1)=0
\ee
 \noindent this multiplet is the $F(4)$ supergravity multiplet.
 Its lowest component, corresponding to  the dilaton in $AdS_6$ supergravity multiplet,
 is a scalar with $E_0=3$ $(m^2=-6)$ and $I=0$.\\
 The tower is obtained as follows
\be \label{magra}G_{q+2}(W)=W^1\overline{W}^1(W^1)^q \ee
 \noindent where the massive graviton, described in eq..
 (\ref{magra}) has $E_0=5+\frac{3}{2}q$ and $I=\frac{q}{2}$.\\
 Note that the $G_{q+2}$ polynomial, although not $G$-analytic,
 satisfies the constraint
\be
D^1_{\alpha}D^{1\alpha}G_{q+2}(W)=0 \ee
 \noindent so that it corresponds to a short representation with
 quantized dimensions and highest weight given by
 $D(E_0=3+3I,0,0;I=\frac{q}{2})$.\\
 We call these multiplets, following \cite{fanta2}, "intermediate
 short" because, although they have some missing states, they are
 not BPS in the sense of supersymmetry. In fact they do not form a
 ring under multiplication.\\
It is worthwhile to mention that the towers given by (\ref{wp}), (\ref{magra}) correspond to the two
 isolated series of UIR's of the $F(4)$ superalgebra argued to exist in
 \cite{min}.\\
There are also long spin 2 multiplets containing $2^8$ state where $E_0$ is not quantized and satisfies the bound $E_0\geq 6$.\\
  Finally let us make some comments on the role played by the flavour symmetry $\cG$.\\
It is clear that,
 since the supersingleton $W^1$ is in a representation of $\cG$ (other than the gauge group of the world-volume theory),
 the $I_p$ and $G_{q+2}$ polynomials will appear in the $p$-fold and
 $(q+2)$-fold tensor product representations of the $\cG$ group.
 This representation is in general reducible, however the 1/2 BPS
 states must have a first component totally symmetric in the $SU(2)$ indices  and, therefore, only
 certains $\cG$ representations survive.\\ Moreover in the $(W^1)^2$ multiplet,
 corresponding to the massless $\cG$- gauge vector multiplets in $AdS_6$, we must pick up the adjoint  representation
 $Adj\cG$ and in $W^1\overline{W}^1$, corresponding to the graviton multiplet, we must pick up  the $\cG$  singlet representation.\\
 However in principle there can be representations in the higher
 symmetric and antisymmetric products, and the conformal field theory should
  tell us which products remains, since the flavor symmetry
  depends on the specific dynamical model.\\

  The states discussed in this paper are expected to appear \cite{oz}, \cite{clp} in the
  K-K analysis of IIA massive supergravity on warped $AdS_6\otimes S^4$. It is amusing that superconformal
 field theory largely predicts the spectrum just from symmetry cosiderations.
  What is new in the $F(4)$ theory is the fact that, since it is
  not a theory with maximal symmetry, it allows in principle some
  rich dynamics and more classes of short representations than the
  usual compactification on spheres.\\
The K-K reduction is related to the horizon geometry of the $D4$ branes in a $D8$ brane background in presence of $D0$ branes \cite{fkpz}, \cite{oz}.\\
Conformal theories at fixed points of $5d$ gauge theories exist \cite{smi}
which exibit global symmetries $E_{N_f+1}\supset SO(2N_f)\otimes U(1)$,  where $N_f\geq 1$ is the number of flavors
 ($N_f$ $D8$ branes) and $U(1)$ is the ``instantons charge'' (dual to the $D0$ brane charge).\\
The $E$ exceptional series therefore unifies perturbative and non perturbative series of the gauge theory.\\
It is natural to conjecture that a conformal fixed point $5d$ theory  can be described by a singleton supermultiplet
 in the fundamental rep. of $E_{N_f+1}$. For the exceptional groups $N_f\geq 5$ these are the {\bf 27} of $E_6$,
the {\bf 56} of $E_7$ and the {\bf 248} of $E_8$ which are respectively complex, pseudo-real and real. The $E_7$ case was considered in ref. \cite{fkpz}. States coming from wrapped $D8$ branes will carry a non trivial representation of $SO(2N_f)$, which, together with some other states, must complete representations of $E_{N_f+1}$. It is possible that from the knowledge of $SO(2N_f)$ quantum numbers of supergravity in $D4-D8$ background one may infer the spectrum of $E_{N_f+1}$ representations and then to realize these states in terms of boundary composite conformal operators.

\section{Acknowledgements}
The authors are grateful to A. Ceresole for her important aid in
the early part of our work.We also thank G.G. Dall'Agata and
especially to A. Zaffaroni for interesting discussions and
suggestions and  P. Fr\'e for his aid in the symbolic
manipulation of the Fierz identities with Mathematica. One of us
(R.D'A.) thanks the theoretical Division of CERN for the kind
hospitality extended to him when most of this work was
performed.\\ The work of R. D'Auria and S. Vaul\'a has been
supported by EEC under TMR contract ERBFMRX-CT96-0045, the work
of S. Ferrara has been supported by the EEC TMR programme
ERBFMRX-CT96-0045 (Laboratori Nazionali di Frascati, INFN) and by
DOE grant DE-FG03-91ER40662, Task C.


\begin{thebibliography}{100}
\bibitem{malda} Juan M. Maldacena, Adv.Theor.Math.Phys. {\bf 2} (1998) 231, Int.J.Theor.Phys. {\bf 38}
(1999) 1113, hep-th/9711200; S. S. Gubser, I. R. Klebanov, A. M.
Polyakov, Phys.Lett. {\bf B428} (1998) 105, hep-th/9802109;  E.
Witten, Adv.Theor.Math.Phys. {\bf 2} (1998) 253, hep-th/9802150
\bibitem{rass} O. Aharony, S. S. Gubser, J. Maldacena, H. Ooguri and Y.
Oz, Phys.Rept. {\bf 323} (2000) 183, hep-th/9905111
\bibitem{nahm} M. Scheunert, W. Nahm and V. Rittenberg, J. Math. Phys. {\bf 17} (1976) 1640; W. Nahm, Nucl. Phys. {\bf B135} (1978) 149
\bibitem{bagu} I. Bars and M. G\"{u}naydin, Comm. Math. Phys. {\bf 91} (1983) 31
\bibitem{fkpz} S. Ferrara, A. Kehagias, H. Partouche, A.
Zaffaroni, Phys.Lett. {\bf B431} (1998) 57, hep-th/9804006
\bibitem{rom} L.J. Romans, Nucl. Phys. {\bf B269} (1986) 691
\bibitem{vn} F. Giani, M. Pernici and P. van Nieuwenhuizen, Phys.
Rev. D {\bf 8} (1984) 1680
\bibitem{oz} A. Brandhuber and Y. Oz,  Phys.Lett. {\bf B460} (1999) 307
\bibitem{clp} M. Cveti\v{c}, H. L\"{u} and C.N. Pope, Phys.Rev.Lett. {\bf 83} (1999)
5226, hep-th/9906221
\bibitem{mtIIA} L. J. Romans Phis. Lett. {\bf B169} (1986) 374
\bibitem{smi} N. Seiberg, Phys. Lett. {\bf B388} (1966) 753,
hep-th/960811; D. R. Morrison and N. Seiberg, Nucl. Phys. {\bf
B483} (1997) 229, hep-th/9609071; P. Intriligator, D. R. Morrison and N. Seiberg, Nucl. Phys. {\bf
B497} (1997) 56, hep-th/9702198
\bibitem{min} S. M. Minwalla, Theor. Math. Phys. {\bf 2} (1998)
781
\bibitem{fanta1} A. Galperin, E. Ivanov, S. Kalitzin, V.
Ogievetsky and E. Sokatchev, Class. Quant. Grav. {\bf 1} (1984)
469
\bibitem{bible} L. Castellani, R. D'Auria and P. Fr\'e,
Supergravity and Superstrings Vol 2 pag 794, World Scientific
(1991)
\bibitem{anna} A. Ceresole and G.G. Dall'Agata, hep-th/0004111
\bibitem{forth} L. Andrianopoli, R. D'Auria and S. Vaul\'a, in
preparation.
\bibitem{breit} P. Breitenlohner and D. Z. Freedman, Phis. Lett. {\bf B115} (1982) 197; Ann. of Phys. {\bf 144} (1982) 249; G. W. Gibbons, C. M. Hull and N. P. Warner, Nucl. Phys. {\bf B218} (1983) 173; W. Boucher, Nucl. Phys. {\bf B242} (1984) 282
 \bibitem{fema} S. Ferrara and L. Maiani, Proc. V Silarg Symposium (World Scientific, Singapore) (1986); S. Cecotti, L. Girardello and M. Porrati, Nucl. Phys. {\bf B268}(1986) 295; A. Ceresole, R. D'Auria, S. Ferrara, P. Fr\'e and E. Maina, Phys. Lett. {\bf B268} (1986) 317
\bibitem{f} W. Siegel, Int. Jou. Math. Phys. {\bf A4} (1989) 2015
\bibitem{g} E. Angelopoulos and M. Laoues, Rev. Math. Phys. {\bf 10}
(1998) 271, hep-th/9806100
\bibitem{h} S. Ferrara, C. Fronsdal, hep-th/0006009
\bibitem{flafro} M. Flato and G. Fronsdal, Lett. Math. Phys. {\bf 2} (1978) 421; Phys. Lett. {\bf 97B} (1980) 236; J. Math. Phys. {\bf 22} (1981) 1100; Phys. Lett. {\bf B172} (1986) 412
\bibitem{fesoca} S. Ferrara and E. Sokatchev, hep-th/000515
\bibitem{hh} G. G. Hartwell and P. S. Howe, Class. Quant. Grav. {\bf 12} (1995) 1823; Int. J. Mod. Phys. {\bf 10} (1995) 3901
\bibitem{fanta2} S. Ferrara and E. Sokatchev, hep-th/0001178
\end{thebibliography}
\end{document}